\documentclass{aa}

\usepackage{natbib}
\usepackage{graphicx}
\usepackage{times}
\usepackage{amsmath}
\usepackage{txfonts}
\newcommand{\msun}{{\,\rm M}_{\odot}}

\newcommand{\kms}{\,{\rm km.s}^{-1}}

\newcommand{\nht}{\ifmmode {{\rm NH}_3} \else {NH{\bas 3}} \fi}
\newcommand{\tco}{\ifmmode {^{13}{\rm CO}} \else {$^{13}{\rm CO}$}\fi}
\newcommand{\dco}{\ifmmode {^{12}{\rm CO}} \else {$^{12}{\rm CO}$}\fi}
\newcommand{\cdo}{\ifmmode {{\rm C}^{18}{\rm O}} \else {${\rm C}^{18}{\rm O}$}\fi}

\newcommand{\htco}{\ifmmode {{\rm H}^{13}{\rm CO}^{+} } \else {${\rm H}^{13}
{\rm CO}^{+}$ }\fi}
\newcommand{\hco}{\ifmmode {{\rm H}^{12}{\rm CO}^{+} } \else {${\rm H}^{12}
{\rm CO}^{+}$ }\fi}
\newcommand{\juz}{\ifmmode {{\rm J}=1\rightarrow\0} \else
{J=1$\rightarrow$0}\fi}
\newcommand{\jdu}{\ifmmode {{\rm J}=2\rightarrow\1} \else
{J=2$\rightarrow$1}\fi}
\newcommand{\jtd}{\ifmmode {{\rm J}=3\!\rightarrow\!2} \else
{${\rm J}=3\!\rightarrow\!2$} \fi}
\newcommand{\jcq}{\ifmmode {{\rm J}=5\!\rightarrow\!4} \else
{${\rm J}=5\!\rightarrow\!4$} \fi}
\newcommand{\as}{\ifmmode {^{\scriptscriptstyle\prime\prime}}
        \else $^{\scriptscriptstyle\prime\prime}$\fi}
\newcommand{\am}{\ifmmode {^{\scriptscriptstyle\prime}}
        \else $^{\scriptscriptstyle\prime}$\fi}
\begin{document}
%
\title{Sub-arcsec imaging of the AB Aur molecular disk and envelope at millimeter
wavelengths: a non Keplerian disk. \thanks{Based on observations carried out with the IRAM Plateau
de Bure Interferometer. IRAM is supported by INSU/CNRS (France), MPG (Germany) and IGN (Spain).} }
\author{Vincent Pi\'etu,  \inst{1} \and St\'ephane Guilloteau \inst{2} \and Anne Dutrey \inst{2}}
\offprints{S.Guilloteau, \email{guilloteau@obs.u-bordeaux1.fr}}
\institute{LAOG, Observatoire de Grenoble, BP 53, F-38041 Grenoble Cedex 9, France \and L3AB, Observatoire de
Bordeaux, 2 rue de l'Observatoire, BP 89, F-33270 Floirac, France}
\date{Received 23 September 2004 /  Accepted 31 March 2005}
\authorrunning{Pi\'etu, Guilloteau \& Dutrey}
\titlerunning{Sub-arcsecond images of AB Auriga}

\abstract{ We present sub-arcsecond images of AB Auriga obtained with the IRAM Plateau de Bure
interferometer in the isotopologues of CO, and in continuum at 3 and 1.3 mm.  These observations
allow us to trace the structure of the circumstellar material of \object{AB Aur} in regions where
optical and IR imaging is impossible because of the emission from the star. These images   reveal
that the environment of AB Aur is widely different from the proto-planetary disks that   surround T
Tauri stars like DM Tau and LkCa15 or HAeBe stars like MWC 480, in several aspects.  Instead of
being centrally peaked, the continuum emission is dominated by a bright, asymmetric (spiral-like)
feature at about 140 AU from the central star.   Little emission is associated with the star
itself.  The molecular emission shows that AB Aur is surrounded by a very extended flattened
structure (``disk''), which is rotating around the star. Bright molecular emission is also found
towards the continuum ``spiral''. The large scale molecular structure suggests the AB Aur disk is
inclined between 23 and 43 degrees, but the strong asymmetry of  the continuum and molecular
emission prevents an accurate determination of the inclination of the inner parts. Analysis of the
emission in terms of a Keplerian disk provides a reasonable fit to the data, but fails to give a
consistent picture because the inclinations determined from \dco~\jdu, \tco~\jdu, \tco~\juz~ and
\cdo~\juz~ do not agree. The mass predicted for the central star in such Keplerian models is in the
range 0.9 to 1.2 $\msun$, much smaller than the expected 2.2 $\msun$ from the spectral type of AB
Aur. Better and more consistent fits to the \tco~\jdu, \tco~\juz~ data are obtained by relaxing the
Keplerian hypothesis. We find significant non-Keplerian motion, with a best fit exponent for the
rotation velocity law of  $0.41 \pm0.01$, but no evidence for radial motions. The disk has an inner
hole about 70 AU in radius. The disk is warm and shows no evidence of depletion of CO. The dust
properties suggest the dust is less evolved than in typical T Tauri disks. Both the spiral-like
feature and the departure from purely Keplerian motions indicates the AB Aur disk is not in
quasi-equilibrium. Disk self-gravity is insufficient to create the perturbation. This behavior may
be related either to an early phase of star formation in which the Keplerian regime is not yet
fully established and/or to a disturbance of yet unknown origin. An alternate, but unproven,
possibility is that of a low mass companion located about 40 AU from AB Aur.}

\maketitle{}



\section{Introduction}

The existence of disks around Herbig Ae/Be stars is now widely accepted but the physical properties
of these disks are poorly known. Contrary to TTauri disks, there are only a few examples of large
Keplerian disks around Herbig Ae stars: the A4 star MWC480 \citep{Mannings_etal1997,
Simon_etal2000}, and the A0 star HD\,34282 \citep{Pietu_etal2003}.

AB Auriga is one of the nearest, brightest and best studied Herbig Ae stars. It has a spectral type
A0-A1 \citep{Hernandez_etal2004} and is located at a distance $D=144\pm^{23}_{17}$~pc following
Hipparcos measurements \citep{VandenAncker_etal1998}. NIR imaging with the HST-STIS
\citep{Grady_etal1999} reveal a large envelope surrounding the star and scattering the stellar
light. This flattened reflection nebulae is seen close to pole-on, up to $r \sim 1300$~AU from the
star. New NIR observations, performed with the Subaru telescope using the Coronographic Imager and
Adaptive Optics systems \citep{Fukagawa_etal2004}, show that the circumstellar matter presents a
spiral structure. AB Auriga has been also observed in MID-IR spectroscopy with ISO
\citep{Meeus_etal2001, Bouwman_etal2000}. These data constrain the dust content several tens of AUs
from the star. The modelling of the SED in this frequency range also shows that the star belongs to
the Group I \citep{Meeus_etal2001}. In the classification of Herbig Ae circumstellar matter, stars
of Group I are surrounded by a flaring disk responsible for the bump in the MID-IR. The disk was
imaged at 11 and 18$\mu$m by \citet{Chen_Jura2003}, using the Keck I telescope. They found the disk
is resolved at 18.7$\mu$m with an approximate diameter of $\sim 1.2''$ or 170~AU. Technical
progresses achieved on optical interferometers allow the first quantitative studies of the very
inner disks. \citet{Monnier_MillanGabet2002} have shown that in such objects, the very inner dust
disk is truncated by sublimation of the dust at temperature around $\sim 1500$~K. For AB Auriga
which is a star of $T_\mathrm{eff} \simeq 10\,000$K, this happens at $r_{in} \sim 0.2-0.3$~AU.
Finally, the surrounding material was also observed at millimeter wavelengths by
\citet{Mannings_Sargent1997} with OVRO, but the derived inclination ($\simeq 75^\circ$) is much
larger than that found from all other (optical or IR) tracers ($\simeq 20-30^\circ$).

Following our successful study of DM Tau in CO isotopologues \citep{Dartois_etal2003}, we decided
to perform a similar study of the AB Auriga environment. Therefore, we observed AB Auriga with the
IRAM interferometer in \dco~\jdu, \tco~\juz~and~\jdu, and \cdo~\juz. We report here the results of
this multi-line, CO isotope analysis. Our goals are double: we want to 1) characterize the physical
structure of the circumstellar matter (vertical temperature gradient, density, kinematics) and 2)
compare quantitatively the large scale mm properties with those of TTauri disks. The observations
and the results are described in Sections 2 \& 3, respectively. Section 4 presents the best model
and we discuss in Section 5, the physical implications.
\begin{figure*}[!ht]
  \resizebox{18.0cm}{!}{\includegraphics[angle=270]{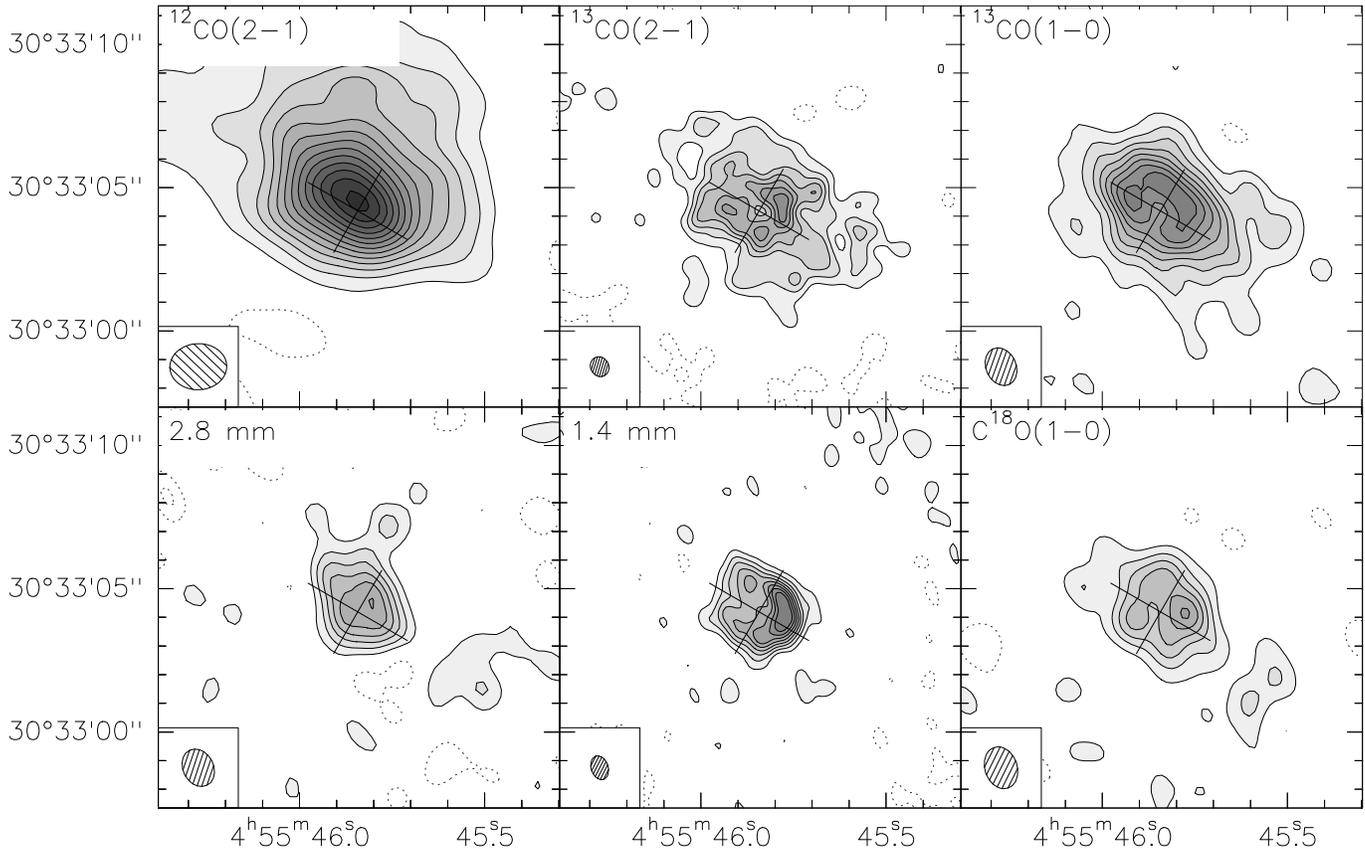}} 
  \caption{Line emission (integrated over the [3.8,8.2] $\kms$ velocity
  range, and continuum emission towards AB Aur. The synthesized beam is indicated in each panel.
  The cross indicates the direction of the major and minor axis of
  the AB Aur disk, as derived from the analysis of the line data,
  but using an inclination of $30^\circ$ for the aspect ratio of the cross.
  Top row, left: Integrated intensity of the \dco\,\jdu\ line. The angular resolution is
  $2.0 \times 1.6''$ at PA~$92^\circ$ and the contour spacing is 0.85 Jy/beam$\kms$ or 12
  $\sigma$.
  Middle: Integrated intensity of the \tco\,\jdu\ line. The angular resolution is
  $0.72 \times 0.63''$ at PA~$29^\circ$ and the contour spacing is 85 mJy/beam$\kms$ or 2.2
  $\sigma$.
  Right: Integrated intensity of the \tco\,\juz\ line. The angular resolution is
  $1.37 \times 1.05''$ at PA~$25^\circ$ and the contour spacing is 34 mJy/beam$\kms$ or 2.3
  $\sigma$.
  Bottom row, left: continuum emission at 2.8 mm. The angular resolution is
  $1.37 \times 1.07''$ at PA~$27^\circ$ and the contour spacing is 0.4 mJy/beam, 28 mK, or
  2.0 $\sigma$. Middle: continuum emission at 1.4 mm. The angular resolution is
  $0.85 \times 0.59''$ at PA~$18^\circ$ and the contour spacing is 1 mJy/beam, 51 mK, or
  2.7 $\sigma$.
  Right: Integrated intensity of the \cdo~\juz\ line. The angular resolution is
  $1.37 \times 1.05''$ at PA~$25^\circ$ and the contour spacing is 25 mJy/beam$\kms$ or 2.1 $\sigma$.
  Coordinates are J2000.0 Right Ascension and Declination.}
  \label{fig:mean}
\end{figure*}
\section{Observations and Results}
\subsection{PdBI data}

The \dco\ observations used 5 antennas and were carried out in winter 2001/2002 in D and C2
configurations. Baselines up to 170\,m provided $2.00 \times 1.60''$ resolution at PA $92^\circ$
for the 1.3\,mm continuum data. We observed simultaneously at 89.2 GHz (HCO$^+$\,\juz, which will
be discussed in a forthcoming paper) and 230.5 GHz (\dco\,\jdu). At 1.3\,mm, the tuning was
double-side-band (DSB) while at 3.4\,mm, the tuning was purely single-side band (LSB). The backend
was a correlator with one band of 10 MHz (spectral resolution 0.23 $\kms$) centered on the
HCO$^+$~\juz\ line, one band of 20 MHz (0.18 $\kms$ resolution) centered on the \dco~\jdu\ line,
and 2 bands of 160 MHz for the 1.3 mm and 3.4 mm continuum, respectively. The phase and flux
calibrators were 0415+379 and 0528+134. The rms phase noise was 8$\degr$ to 25$\degr$ and 15$\degr$
to 50$\degr$ at 3.4 mm and 1.3 mm, respectively, which introduced position errors of $\leq
0.1\arcsec$, and a seeing better than $0.3''$. The observation time was shared with other sources:
MWC480 and LkCa15 or CQ Tau and MWC758, AB Aur using only a small fraction of the transit time. The
total on source integration time is $\sim 10$ hours. As a consequence, the amplitude and phase
calibration of these sources is homogenous and the flux density of all sources were estimated
simultaneously. This allows us to make accurate comparisons of the spectral index of the continuum
emission. In all  cases, the flux density scale was referred to MWC349, for which we used a flux of
$S(\nu) = 0.95 \mathrm{Jy} (\nu/\mathrm{100~GHz})^{0.6}$.

The \cdo\,\juz, \tco\,\juz\ and  \jdu\ observations were performed between 2001 and 2004.
Configurations D, C2, B1 (5 antennas) and A (6 antennas) were used, and provided baselines up to
400~m. This provides angular resolution of $0.80 \times 0.55''$ at PA $39^\circ$ at 220 GHz, with a
residual seeing $< 0.2''$ and position errors $\leq 0.05\arcsec$ after calibration. The correlator
provided a spectral resolution of 0.09 $\kms$ for \tco\,\jdu\ and 0.18$\kms$ for the \tco\,\juz\
and \cdo\,\juz\ lines. The same phase and flux calibrators were used. The total integration time is
around 30 hours, half of which was spent on the long baseline (A configuration) observations.

We used the GILDAS software package to reduce the data. Images are presented at different angular
resolutions, obtained by applying a taper and re-weighting the data when needed. However, the
natural weights were used in the disk modelling.

\subsection{30-m data}

In 2000, we also obtained a single-dish spectrum of \tco\,\jdu\ at the 30-m telescope,
which when compared to the PdB measurements indicates at most 20 \% of the flux was missed
by the interferometer. Accordingly, it is unlikely that the \tco\ measurements are
significantly affected by missing flux, specially within the inner 10 to 20$''$. However,
the \dco\,\jdu\ line is much brighter and most likely hampered by structures within a more
extended region.

\begin{figure*}
  \resizebox{15.0cm}{!}{\includegraphics[angle=0]{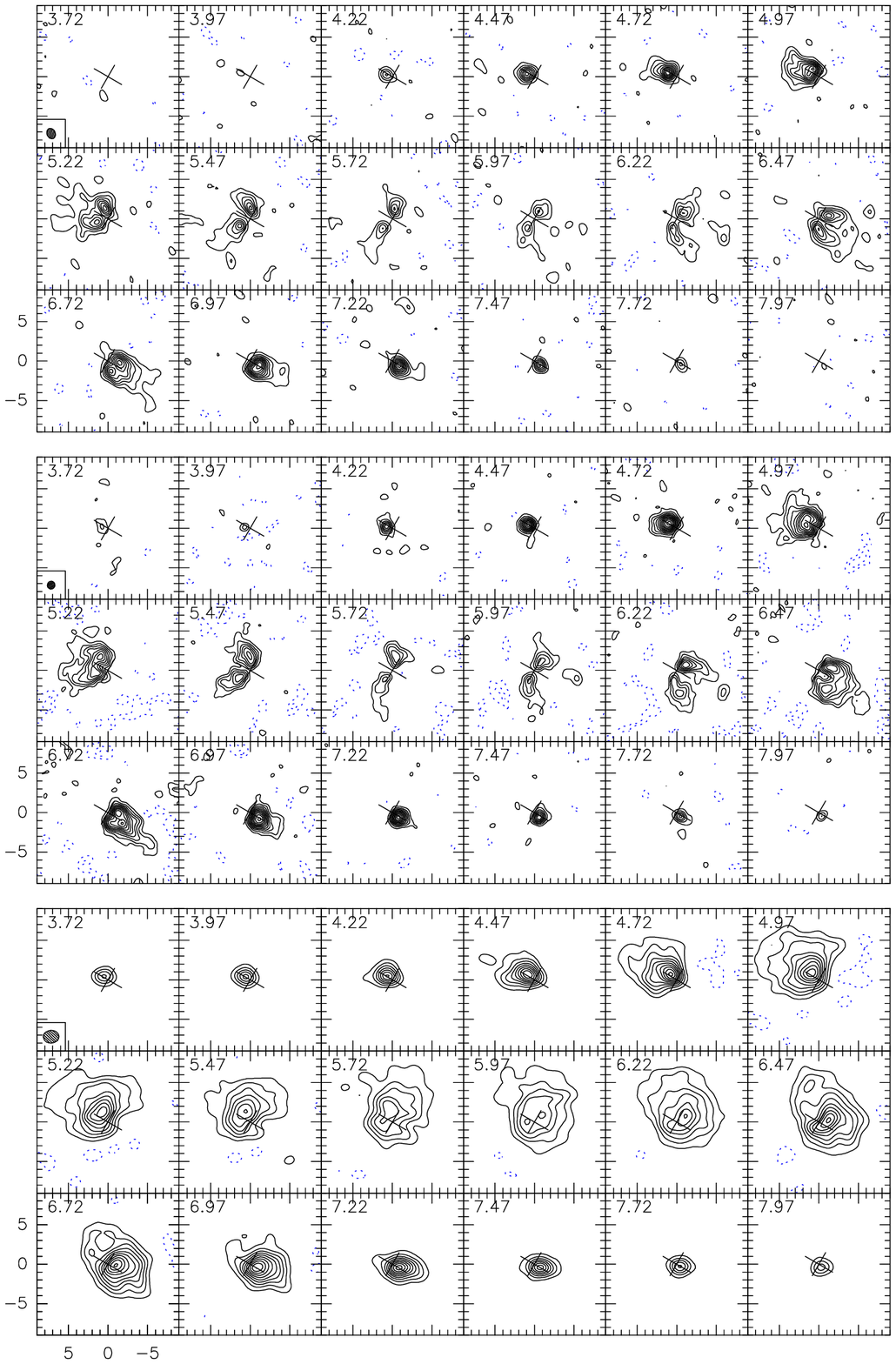}} 
  \caption{Channel maps of the CO isotopologues in AB Aur.
  The LSR velocity, in km.s$^{-1}$, is indicated in the upper left corner
  of each panel. Coordinates are in arsec from the star position.
  All data have been smoothed to the same spectral resolution, 0.25 $\kms$.
  Top: \tco\,\juz\ line, spatial resolution $1.37 \times 1.05''$ at PA~$28^\circ$, contour
  spacing 30 mJy/beam, or 2.1~K, or 2.2 $\sigma$.
  Middle: \tco\,\jdu\ line, spatial resolution $1.00 \times 0.95''$ at PA~$168^\circ$, contour
  spacing 0.1 Jy/beam, or 2.6~K, or 2.6 $\sigma$.
  Bottom:  \dco\,\jdu\ line, spatial resolution $2.0 \times 1.6''$ at PA~$92^\circ$, contour
  spacing 0.5 Jy/beam, or 3.6~K, or 7 $\sigma$. Negative contours are dashed, and the zero contour is omitted.}
  \label{fig:maps}
\end{figure*}

\section{Results}

Figure \ref{fig:mean} is a montage displaying high resolutions images of the continuum emission at 2.8 mm and
1.4 mm, and of the integrated line emission of \dco\,\jdu\ \tco\,\jdu\ and \tco\,\juz\ transitions. The 1.4 mm
emission was obtained by merging the 220 and 230 GHz data in together to improve signal to noise; the flux was
scaled to a reference frequency of 220 GHz assuming a spectral index of 3.0 for the emission. The (lower
resolution) emission at 110 GHz is also presented. The total continuum flux at 1.4\,mm is $85 \pm 5$~mJy, and at
2.8 mm, $11 \pm 2$ mJy.

Figure \ref{fig:maps} presents the channels maps of the \dco\,\jdu, \tco\,\jdu\ and \tco\,\juz\
transitions.  The \tco\,\jdu\ data has been smoothed to about 1$''$ resolution to help the
comparison with the \juz\ transition.

\begin{figure}[ht]
  \resizebox{\columnwidth}{!}{\includegraphics[angle=270]{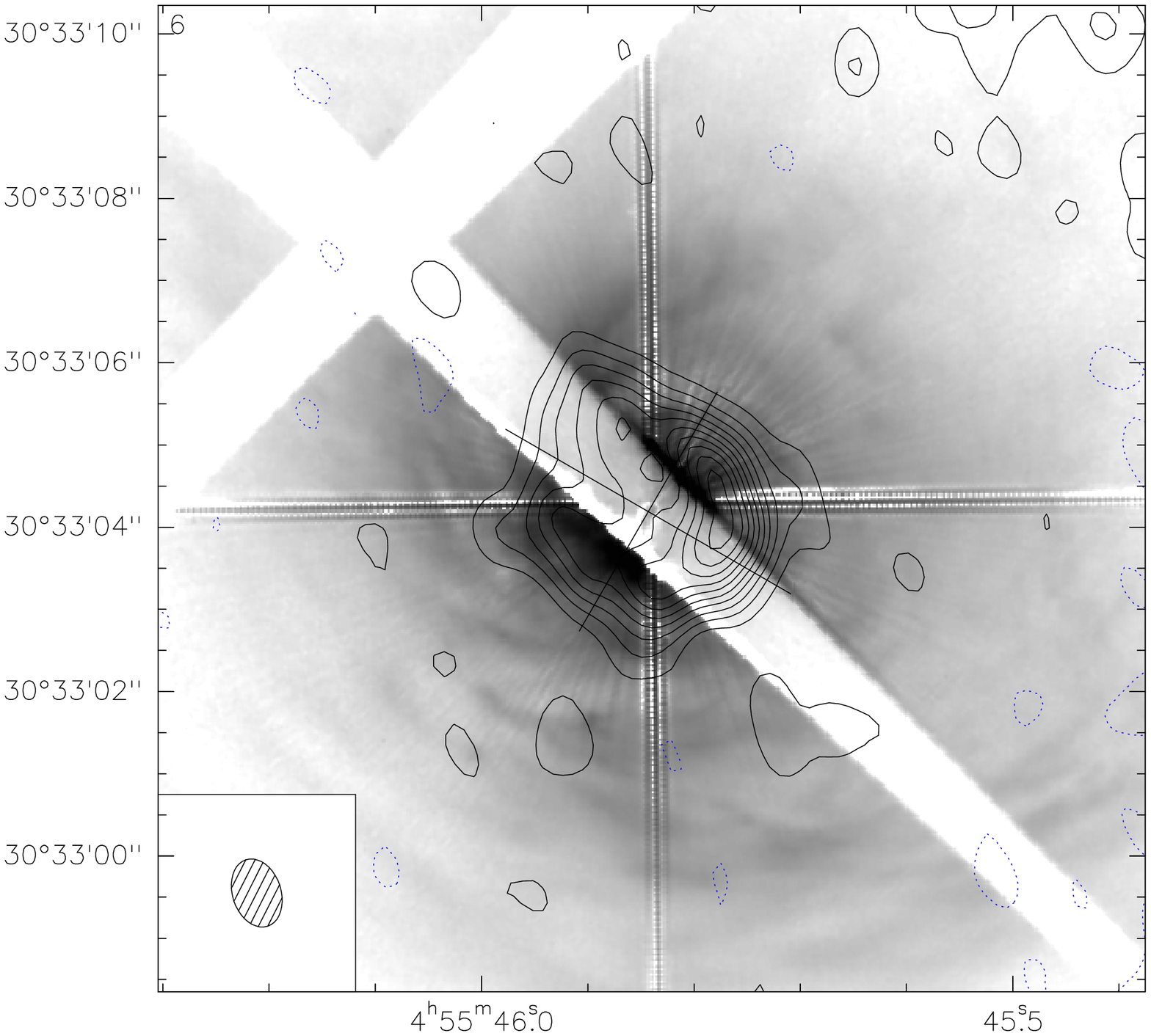}} 
  \caption{1.4 mm continuum data (in contours) superimposed on the HST image from \citet{Grady_etal1999}, in
  grey scale. The angular resolution is $0.85 \times 0.59''$ at PA~$18^\circ$. The contour spacing is 0.9 mJy/beam,
  corresponding to 45 mK, or 2.4 $\sigma$, with the zero contour omitted. The cross is as in Fig.1}
  \label{fig:hst}
\end{figure}

In Figure \ref{fig:hst}, the continuum emission is super-imposed on the STIS image from
\citet{Grady_etal1999}. No position adjustment has been made in doing so: the absolute astrometric
accuracy of the PdB measurement is better than $\simeq 0.05''$. The best fit position of the
centroid of the \tco~ emission is $\alpha=$~04:55:45.843, $\delta=$~30:33:04.21 only 0.03$''$ from
the extrapolated Hipparcos position. The HST STIS image (which was centered on the Hipparcos
coordinates for 1999) was shifted southward by 0.08$''$ to correct for the proper motion of AB Aur
(taken from the Hipparcos catalog) between the epochs of the two observations (1999 and 2003). The
emission is clearly concentrated in an arc-like structure about 1$''$ from the central star. The
flux density measured at 1.4 mm in the direction of the central star is $\simeq 5 \pm 1$~mJy, but
this should be considered as an upper limit because of contamination from the arc-like structure
due to the limited angular resolution.

In Figure \ref{fig:mean}, the line emission was integrated between 3.8 and 8.2 $\kms $. This
interval contains most of the line profile, but weak emission in the line wings is still detectable
2 $\kms$ away (specially in \dco). The integrated line emission maps reveal a clear inner hole in
the distribution of the emission from the CO isotopes, as well as enhanced emission at the location
of the continuum ring. Note in particular that, while the 2.8\,mm continuum hardly shows the
central depression, the \tco~ and \cdo\,\juz\ line emissions obtained at the same angular
resolution clearly show the ring.


\subsection{Analysis Method}
To better quantify the properties of the AB Aur circumstellar material, we used an approach similar
to that applied to DM Tau by \citet{Dartois_etal2003}.  The analysis is based on a new radiative
transfer code, with non-LTE capabilities, which will be presented in a forthcoming paper
\citep{Pietu_etal2004}. Apart from discretization details, the LTE part (which was used exclusively
here) is however equivalent to the code described in \citet{Dutrey_etal1994}. As in
\cite{Guilloteau_Dutrey1998}, the comparison between models and data and the minimization is done
inside the UV plane, using natural weighting. The physical model is that of a rotating, flared disk
in hydrostatic equilibrium, with power law distributions for the density, temperature, rotation
velocity and scale height as function of radius. Since the AB Aur disk is relatively extended, we
have applied primary beam attenuation to the model images before computing the model visibilities.
We have also improved the minimization technique by searching the minimum of the $\chi^2$ with the
help of a Levenberg-Marquard method, instead of a simple grid search previously. The two methods
were intensively compared, the new one improving the convergence speed by a factor $\sim 20-30$ on
typical data, and providing more accurate estimate of the error bars since it takes into account
the correlation between parameters.
Table \ref{tab:abaur} summarizes the parameters found for the best models for the various lines and
the continuum.

\subsection{Continuum Emission: a spiral pattern?}

The continuum emission from AB Aur is quite peculiar. While most other T Tauri stars and HAeBe
stars observed at high angular resolution show a centrally peaked emission, the emission from AB
Aur is reminiscent from that of GG Tau, except for the angular scale which is a factor of 2 smaller
here. The emission peaks accurately coincide with bright features observed in the STIS image of
\citet{Grady_etal1999}, but our observations allows to trace the regions which are occulted by
coronagraphic mask or by the telescope spider in the HST observations.

\begin{figure}
  \resizebox{\columnwidth}{!}{\includegraphics[angle=270]{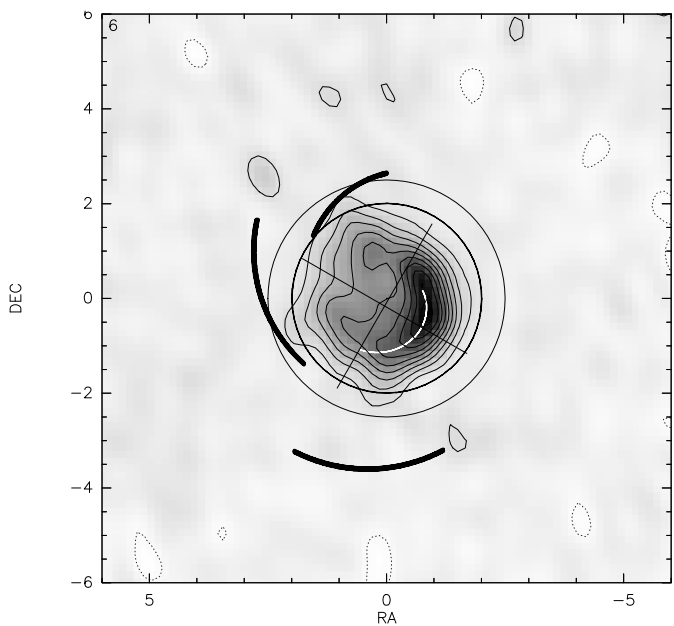}} 
  \caption{Deprojected image (from an inclination of 23$^\circ$) of the continuum emission of AB Aur at 1.4 mm. The black curves
  indicate the approximate location of the spiral arms detected by \citet{Fukagawa_etal2004}.
  The white curve is a possible inward spiral arm revealed by this observation.
  The outer ring, at 350 AU, indicates the measured outer radius of the continuum emission.
  The inner ring, at 280 AU, is the location of the feature observed by \citet{Pantin_Bouwman2005} at
  20.5 $\mu$m.
  }
  \label{fig:spiral}
\end{figure}

No reliable estimate of the inclination of the structure can be obtained from the 1.4 mm image
only, because of the highly asymmetric nature of the emission. We have deprojected the emission
assuming at position angle $-31^\circ$, as derived from our analysis of the \tco\ data, and an
inclination of 23$^\circ$. These values are also in agreement with the numbers quoted by
\citet{Fukagawa_etal2004} from the scattered light in H-band.  The result is presented in
Fig.\ref{fig:spiral}: the asymmetry seen in continuum may follow inwards the spiral-like features
detected in scattered light by \citet{Fukagawa_etal2004}. By extension, we shall refer to this
structure as ``the spiral''.

\begin{figure}
  \resizebox{\columnwidth}{!}{\includegraphics[angle=270]{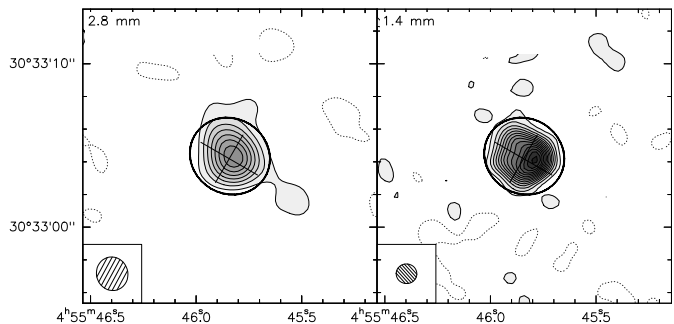}} 
  \caption{Images of the continuum 2.8 and 1.4 mm emission of AB Aur at 1.4 mm.
  1.4 mm image: resolution $1.3''$, contour spacing 1.3 mJy/beam (22 mK, 2 $\sigma$)
  2.8 mm image: resolution $2.0''$, contour spacing 0.6 mJy/beam (16 mK, 2 $\sigma$)
  The ellipse, at 350 AU, indicates the measured outer radius of the continuum emission.
  }
  \label{fig:smooth}
\end{figure}

Although the spiral near 160 AU is the most striking feature in the continuum, weak extended
emission is also found at larger distances from the star. When interpreted in terms of a truncated
disk, the best fit requires a disk with an inner radius of $\simeq 110$~AU, an outer radius $\simeq
350$~AU, and a power  law distribution for the surface density of $p \simeq 2.3$ (see Table
\ref{tab:abaur}). This outer radius is within $2 \sigma$ of the 280 AU ring detected at 20.5 $\mu$m
by \citet{Pantin_Bouwman2005}. Negligible emission arise from beyond 350 AU, as shown by the lower
resolution, higher brightness sensitivity images displayed in Fig.\ref{fig:smooth}.
A larger inner radius is found at 1.4~mm than in the CO lines (see Table 1). This is presumably due
to a combination of the smaller dust opacity with a progressive fall of the surface density inside
the ring. When modelled with a sharp edge, this naturally results in the most optically thin
tracer, namely the dust emission, having the largest apparent inner radius.

A simultaneous fit of the 2.7 and 1.4 mm continuum data allows to derive the spectral index of the
dust emissivity, $\beta = 1.4\pm 0.2$, assuming the emissivity $\kappa(\nu) = 0.1
\mathrm{cm}^2.\mathrm{g}^{-1}({\nu}/10^{12}\,\mathrm{Hz})^{\beta}$.

\subsection{Line Emission: the Disk}
\label{sub:line}
The double peak profiles, combined with the strong velocity gradient along the
major axis of the emission which is found in all transitions, suggest that the material around AB
Aur is dominated by a circumstellar disk in Keplerian rotation.

We thus attempted to model the emission by a simple Keplerian disk model, in hydrostatic
equilibrium, as performed successfully for T Tauri stars like DM Tau \citep{Guilloteau_Dutrey1998},
GM Aur \citep{Dutrey_etal1998} and other PMS stars including the Herbig Ae star MWC 480
\citep{Simon_etal2000}. We modelled the circumstellar matter, gas and dust, using the same strategy
than those applied in \citet{Dartois_etal2003} for the DM Tau disk. The \dco~\jdu~ transition
traces  the CO surface of the disk while the \tco~\juz~ \& \jdu~ lines trace material closer to the
mid-plane.
\begin{table*}[!ht]
 \caption{Best parameters for the AB Aurigae disk derived from $\chi^2$ minimization in the $UV$ plane.}\label{tab:abaur}
\begin{tabular}{|l|rll|rl|rl|rl|}
 \hline
 Assumed~ Distance & D (pc)~ = & \multicolumn{8}{c|}{140} \\
 \hline
 \hline
  & \multicolumn{3}{|c|}{ }& \multicolumn{2}{|c|}{ }& \multicolumn{2}{|c|}{ }& \multicolumn{2}{|c|}{ }\\
  Lines & \multicolumn{3}{|c}{\dco~\jdu ~~ \dag} & \multicolumn{2}{|c|}{\tco~\jdu ~~ \ddag}&
  \multicolumn{2}{|c|}{\tco~\juz ~~ \ddag}& \multicolumn{2}{|c|}{\cdo~\juz ~~ $\sharp$}\\
  & \multicolumn{3}{|c|}{ }& \multicolumn{2}{|c|}{ }& \multicolumn{2}{|c|}{ }& \multicolumn{2}{|c|}{ }\\
 \hline
  & \multicolumn{3}{|c|}{ }& \multicolumn{2}{|c|}{ }& \multicolumn{2}{|c|}{ }& \multicolumn{2}{|c|}{ }\\
 Systemic velocity & $\mathrm{V}_\mathrm{LSR}$ (km.s$^{-1}$)~= & 5.89 & $\pm$ 0.01 & 5.84 & $\pm$ 0.01 & 5.87 & $\pm$ 0.03 & 5.84 & $\pm0.02$\\
 Orientation & PA~($^{\circ}$) = & -27.5 & $\pm 0.5$ & -31.3 & $\pm 0.3 $ & -30 & $\pm 1 $ & -29 & $\pm 1$ \\
 Inclination & $i$~($^{\circ}$) = & 33 & $\pm 1$ & 42 & $\pm 1 $ & 39 & $\pm 2 $ & 36 & $\pm 3$ \\
 Inner radius & $R_\mathrm{in} $ (AU)~ = & 45 & $\pm 3$    & 72 & $\pm$ 5 & 77 & $\pm 5$ & 67 & $\pm 6$ \\
 Outer radius & $R_\mathrm{out}$ (AU)~ = & 1050 & $\pm$ 10  & 890  &$\pm$ 10 & 1300 & $\pm 100$ & 600 & $\pm 60$ \\
 Turbulent linewidth  & $\Delta \mathrm{v}$ (km.s$^{-1}$)~$=$ &0.38 &$\pm$ 0.02 & 0.22 &$\pm 0.02$ &0.26 &$\pm 0.02$ & 0.18 & $\pm 0.05$\\
  & \multicolumn{3}{|c|}{ }& \multicolumn{2}{|c|}{ }& \multicolumn{2}{|c|}{ }& \multicolumn{2}{|c|}{ }\\
 \hline
 \multicolumn{10}{|c|}{ } \\
 \multicolumn{10}{|c|}{Molecular Column density law:~~ $\Sigma(r) = \Sigma_{100} (\frac{r}{100 \rm{AU}})^{-p}$} \\
 Column Density      & & & & & & & & &\\
 ~~~~ at 100 AU      & $\Sigma_{100}$ (cm$^{-2}$)~ = & \multicolumn{2}{c|}{($1.4~10^{19}$) } & $2.3~10^{17}$ &$\pm 0.3~10^{17}$
                     & $3.2~10^{17}$& $\pm0.8~10^{17}$ & $9.5~10^{16}$ & $\pm 1.3~10^{16}$ \\
 ~~~~ exponent       & $p$~=  & \multicolumn{2}{c|}{(2.5)} & 2.4 &$\pm 0.1$ &2.5 & $\pm 0.3$ & \multicolumn{2}{c|}{3 $\sharp$} \\
 \hline
 \multicolumn{10}{|c|}{ } \\
 \multicolumn{10}{|c|}{Temperature law:~~~ $T(r)~ = T_{100} (\frac{r}{100\,\rm{AU}})^{-q}$ } \\
 Temperature     &  & & & & & & & & \\
 ~~~~ at 100 AU  & $T_{100}$ (K)~ = & 68 &$\pm 1$  & 38 & $\pm$ 2 & 34 & $\pm 2$ & 20 & $\pm 3$ \\
 ~~~~ exponent   & $q=$ & 0.77 &$\pm 0.02$ & $0.16$ & $\pm 0.03$ & 0.10 & $\pm 0.14$ & -0.6 & $\pm 0.2$ \\
 \hline
 \multicolumn{10}{|c|}{} \\
 \multicolumn{10}{|c|}{Velocity law:~~~~~~ $\mathrm{V}(r) = \mathrm{V}_{100} (\frac{r}{100\,\rm{AU}})^{-\mathrm{v}}$}\\
 Velocity at 100 AU    & $\mathrm{V}_{100}$ (km.s$^{-1}$)~= & 3.06 & $\pm 0.06$ & 2.54 & $\pm 0.03$ & 2.73 & $\pm 0.12$ & 2.75 & $\pm 0.13$ \\
 Velocity exponent     & $\mathrm{v}$~=        & 0.82 & $\pm 0.02$ & 0.42 & $\pm 0.01$  & 0.37 & $\pm 0.02$ & 0.47 & $\pm 0.04$\\
 \hline \hline
 \end{tabular}
\begin{minipage}{8.0cm}
\begin{tabular}{|l|rll|rl|rl|}
 \hline
 \multicolumn{8}{|c|}{} \\
\multicolumn{8}{|c|}{ Continuum Results \textbf{*}} \\
 \multicolumn{8}{|c|}{} \\
 \multicolumn{8}{|c|}{Dust:~~$\kappa_\nu = \kappa_o\times(\frac{\nu}
 {10^{12}\,\mathrm{Hz}})^{\beta}$} \\
 \multicolumn{8}{|c|}{} \\
 \hline
 Absorption law & $\kappa_o$~(cm$^2$/g)~~= & \multicolumn{6}{c|}{(0.1)} \\
 Dust exponent          & $\beta$~~= & \multicolumn{6}{c|}{$1.4~~\pm 0.2$} \\
 \multicolumn{8}{|c|}{ } \\
 \multicolumn{8}{|c|}{Surface Density law:~~ $\Sigma(r) = \Sigma_{100} (\frac{r}{100\,\rm{AU}})^{- p}$} \\
 \multicolumn{8}{|c|}{} \\
Surface Density & $\Sigma_{100}$~(H$_2$~cm$^{-2}$)~~= & \multicolumn{6}{c|}{$6.3\,10^{23}$~~$\pm 0.8\,10^{23}$} \\
 ~~~~ at 100 AU & $\Sigma_{100}$~(g.cm$^{-2}$)~~= & \multicolumn{6}{c|}{$2.7$~~$\pm 0.3$} \\
 ~~~~ exponent       & $p$~~= & \multicolumn{6}{c|}{$2.3$~~$\pm 0.2$} \\
 Inner radius & $R_i$ (AU)~~= & \multicolumn{6}{c|}{$115$~~$\pm 5~~~$} \\
 Outer radius & $R_d$ (AU)~~= & \multicolumn{6}{c|}{$350$~~$\pm 30~~$} \\
   Mass & $M_d$ ($\msun$)~~= & \multicolumn{6}{c|}{ $1.6~10^{-2}$~~$\pm 0.2~10^{-2}$ } \\ \hline
 \hline
\end{tabular}
\end{minipage}
\hspace{0.5cm}
\begin{minipage}{9.2cm}
~~\ddag\,)~  For the $^{13}$CO lines, the error bars are derived from the 1$\sigma$ formal errors
from the $\chi^2$ fit. We checked that all parameters (except of course $V_{100}$) are essentially
independent of the inclinations over the $25 - 45^\circ$ range (see Section \ref{sub:line}).

~~\dag\,)~  For $^{12}$CO, multiple minima exist. We have selected a solution in which the scale
height is determined from the  $^{13}$CO results: $h_{100} = 12$~AU, $h = -1.40$. The error bars,
when mentioned, represent the typical (1 $\sigma$) variations of the parameters obtained when
varying all other ones.  Numbers in parenthesis represent assumed (fixed) values.

~~$\sharp)$~  For the  C$^{18}$O, only $p-q$ is actually constrained: $p-q = 3.6 \pm 0.2$. See text
for details.

~~\textbf{*}\,)~ For the continuum data,  the temperature was taken as 35~K at 100~AU with an
exponent of $q=0.3$. The surface densities and mass (and their errors) are given for $\beta = 1.4$.
An additional 30 \% uncertainty should be added for these quantities, because of the error on
$\beta$.
\end{minipage}
\end{table*}

Clearly, AB Aur is surrounded by a very large disk. All 4 spectral lines give a consistent value
for the orientation of the disk axis, $-30^\circ \pm 1^\circ$ East from North.  The outer radius is
of order 1000 AU, except for \cdo~\juz\ which is significantly smaller (600 AU). Note that the
\jdu\ transitions, specially from \dco\ may underestimate the true radius because of a lack of
sufficient short spacings.

Both \tco\ transitions give a consistent value for the rotation velocity at 100 AU, $\mathrm{v}
\sin(i) = 1.60 \pm 0.05 \kms$, and a slightly different value from \dco. When attempting to fit the
data with a simple Keplerian disk model, the derived inclinations are different (34, 36 and
43$^{\circ}$), and the derived stellar mass (0.9 -- 1.2 $\msun$) far from the expected value for an
A0 star. From Table 1, it is clear that a simple Keplerian disk model fails to fit properly all
observed transitions, even if account is taken of a possible vertical temperature gradient in the
disk as has been observed in DM Tau by \citet{Dartois_etal2003}. The most significant result is
that the velocity law deviates from Keplerian motions. From \tco~\jdu, the exponent is
$\mathrm{v}=0.42 \pm 0.01$, instead of $\mathrm{v}=0.50$ for Keplerian motions, and the \juz\ gives
a consistent result. The law deduced from \dco\ is much steeper, with an exponent of order 0.8.

It is of course conceivable that the \dco\,\jdu\ data be discarded in such an analysis, because of
the confusion with the surrounding molecular cloud. Indeed, the fit by emission from a disk leaves
significant residual, and cannot represent for example the excess emission in the N-E which is seen
in CO at velocities between 4.7 and 6.8 $\kms$. The large turbulent width found in this analysis is
just a result of attempting to fit non-symmetric features with a pure rotating disk. Despite these
caveats, an important result from the CO data is the high temperature  ($T_{100} \simeq 65 - 70
$~K) required in the inner 100 AU.

The two transitions of the \tco\ isotopologues yield reasonably consistent results. Some of the
differences can be attributed to the importance of the emission coming from the continuum spiral in
the \tco\,\jdu\ data. This bright emission dominates the total flux, and biases the derived
inclination towards higher value, because the spiral curves inwards towards the North-West. If the
aspect ratio of the \tco\,\jdu\ integrated emission is taken as a measure of the inclination of a
simple disk, as did \citet{Mannings_Sargent1997}, quite high values are found ($\simeq 70^\circ$).
Significantly lower values are obtained when the constraints from the kinematics of a rotating disk
is incorporated, but the result on the inclination is still biased because of the emission from the
spiral.
The \tco\,\juz\ line being observed with lower angular resolution is less sensitive to the spiral
feature, and yields an inclination $i = 39 \pm 2^\circ$. Given the departure from circular
symmetry, lower values (e.g. down to 25$^\circ$) cannot be excluded. We note also that it is
necessary to introduce an inner radius of about 70 AU to best reproduce the \tco\ emission (see
Table). This inner gap is clearly visible in the integrated intensity maps of Fig.\ref{fig:mean}.

The most clear result from the \tco\ analysis is that the best fit to the velocity law is  obtained
with an exponent $\mathrm{v}=0.41 \pm 0.01$. This is a very robust result: we checked that the
derived exponent is independent of the assumed inclination over a range from $20^\circ$ up to
$50^\circ$. This robustness also applies to most other parameters (column density law, inner and
outer radii, temperature law, turbulent width). Using a stellar mass $M_* = 2.2 \msun$ for AB Aur,
the Keplerian speed at 100 AU is $\simeq 4.4 \kms$. Accordingly, for the nominal inclination of
$39^\circ$, the orbital velocities are sub-Keplerian ($\mathrm{v}(r) < \sqrt{GM_*/r}$) out to the
disk edge. However, if the inclination is lower than $\simeq 22^\circ$, the orbital velocities are
super-Keplerian outwards of $\simeq 100-150$~AU.


\cdo\,\juz\ data, although more noisy, also give a $i \simeq 39^\circ$ inclination. This result is
important, since \cdo\,\juz\ is the most optically thin of all observed transitions, and thus the
less likely to be affected by an extended envelope. The velocity exponent found from \cdo\,
$\mathrm{v}=0.47\pm0.04$ (which still remains compatible with the \tco~results at $2 \sigma$), is
essentially Keplerian, but the projected velocity at 100 AU is much smaller than expected from an
A0 star, as in the case of other transitions. To reconcile the velocities with the expected stellar
mass would require an inclination of $\simeq 23^\circ$. Note also that, because of its optical
thinness, this transition mostly samples the inner radii with little contribution from the outer
parts of the disk ($r > 300$~AU) and the envelope. Also, an independent constraint on the surface
density and the temperature laws is no longer possible; for an optically thin J=1-0 line, the data
only constrain the ratio $\Sigma / T$ and the value of $p-q$ \citep{Dartois_etal2003}.

For all transitions, we have also checked whether adding a radial velocity component would provide
a better fit to the kinematic pattern. For \tco~ and \cdo~, the upper limit on a radial velocity is
below 0.05 $\kms$. For \dco~, a slightly better fit is found by adding outward motions of order
$0.2 \kms$. The significance of this result should not be overestimated, since the excess emission
towards the North near $\mathrm{V}_\mathrm{LSR} = 6.5 \kms$ (see Fig.\ref{fig:maps}) may bias the
result. Note that we can discriminate between inward and outward motions only because we know the
full 3-D orientation of the AB Aur disk (the Southern part being towards us, in agreement with
\citet{Fukagawa_etal2004}).

\section{Discussion}
The observations presented above bring new information about the circumstellar environment of AB Aur, which
appears to be very different from other proto-planetary disks.

\subsection{The ``spiral''}
First, the sub-arcsecond image of the continuum emission at 1.4 mm extends inwards the spiral
structure detected in the near IR by \citet{Fukagawa_etal2004}. Considering the sense of rotation,
the new portion of the spiral is trailing, as the previous arms. Dust emission at long wavelengths
being proportional to $\kappa_{\nu} \Sigma T$, this enhanced brightness could be due to 3 effects:
1) a column density, 2) a change in dust properties, with larger grains in the spiral, or 3) a
larger temperature. However, grain growth cannot explain the enhanced molecular emission which is
also seen in the spiral. It is not straightforward to disentangle between the two remaining effects
from the present data only. Taken at face value, the analysis of \tco\ lines seem to indicate a
flat (or even rising) temperature throughout the disk. However, this result is to be taken with
care, since the \tco\ lines are optically thick only in the inner spiral, whose structure is not
adequately fitted by the disk model. A better insight is given by the fact that the \cdo\,\juz\
line emission is proportional to $\Sigma / T$ \citep{Dartois_etal2003}, rather than $\Sigma T$ for
optically thin dust emission. Since the brightness contrast in \cdo\,\juz~ is similar to that in
the 1.4 mm continuum, it thus suggests the enhanced brightness is a column density effect.

The difference between the apparent inner radius in dust (110 AU) and in CO ($\simeq 70$~AU from all lines)
indicates that the structure of the inner region is much more complex than a simple hole, in sharp contrast with
the case of GG Tau \citep{Dutrey_etal1994, Guilloteau_etal1999}, but its detailed structure is beyond the
resolving power of these observations.

\subsection{The Inclination Problem and non-Keplerian Velocities}
By revealing non circular structures, these observations offer some explanation for the widely
different inclinations which were derived in the past from mm tracers and optical images. High
inclinations are ruled out by our data. Although an inclination of $\simeq 38^\circ$  seems
adequate to represent most of the emission, it is certainly biased by the non-circularity of the
emission. Given the uncertainties, inclinations as low as 23$^\circ$ cannot be excluded.

On the other end, we note that the inclinations derived for the inner disk ($< 2 $~AU) through
near-IR interferometry by \citet{Eisner_etal2003} could be biased towards low values by
non-circular structure, if the ``spiral'' pattern persists at such distances from the star.

In any case, the orbital velocities significantly differ  from the Keplerian speed. In addition,
the turbulence (see Table 1) appears somewhat larger in AB Aur ($0.2 \kms$  if we exclude \dco)
than in other sources like e.g. DM Tau. Unfortunately, the exact value of the inclination plays a
major role in the understanding of the environment, because depending on the assumption about
projection effects, the orbital velocities of the material surrounding AB Aur can be sub-Keplerian
or super-Keplerian throughout most of the disk.

\subsection{Kinetic Temperature}

In this respect, the AB Auriga disk behaves like those around other T Tauri stars. There is a clear
evidence for a kinetic temperature gradient from the disk mid-plane to the disk surface and
envelope, with the disk plane being cooler (around 30~K, as traced by \tco\ and \cdo) than the
envelope ($\sim 70$~K at 100 AU, as traced by \dco, see Table \ref{tab:abaur}), as expected from
heating by the central star. The temperature remains high ($\simeq 30$~K) throughout the disk, as
expected from the much higher luminosity of AB Aur compared to typical T Tauri stars. However, the
temperature derived from \dco~  falls below that derived from \tco~ lines beyond a radius about 400
-- 500 AU. This could be due to self-absorption by the (colder) envelope.

\subsection{Molecular Abundances and Disk Mass}
\label{sec:mass}
The derived outer radius of the dust disk is much smaller ($\simeq 400$~AU) than that the molecular
disk ($\simeq 1300$~AU). At 200 AU, the ratios of the molecular column densities and H$_2$ surface
density derived from the dust indicate a molecular abundance [\tco/H$_2$] $\simeq 5\,10^{-7}$,
similar to that of the Taurus cloud ($10^{-6}$, \citet{Frerking_etal1982}. The value of the dust
opacity index $\beta = 1.4 \pm 0.2$ (giving $\kappa_\mathrm{220GHz} = 0.012$~cm$^{2}$/g), is at the
high end of the values found in circumstellar disks (and in particular of those observed
simultaneously with AB Aur, which share the same flux calibration, see Section 2.1).
\citet{Dutrey_etal1996} indicate a mean value of $\beta = 0.95 \pm 0.11$ from an ensemble of disks.
This suggests that the disk around AB Aur may be younger than average, and that dust grains have
not evolved as much as in disks around T Tauri stars. The relatively high temperatures (above 30~K)
prevent condensation on dust grains, and probably explain why CO is not significantly depleted,
contrary to most other circumstellar disks (e.g. \citet{Dutrey_etal1996}).

The lack of  detectable dust emission beyond 350 AU could indicate that the dust opacity drops
there. Since the molecular content only changes smoothly with radius, this suggest that the dust
absorption coefficient at 1.4 mm is smaller beyond 350 AU than inside, i.e. that the dust is
significantly less evolved outside. Another alternative is a steepening of the surface density
distribution beyond $\simeq 300$ AU, since the values derived from the dust ($p=2.3$) is slightly
lower than that derived from the CO molecules ($p=2.7$). The latter possibility is however not
supported by the \cdo\ data, which are only sensitive to the inner regions. It is tempting to
relate the change of dust properties near 350 AU to the bright 20 $\mu$m ring detected by
\citet{Bouwman_Pantin2003} at 280 AU.

Both the high $\beta$ value, and the possible dust opacity drop beyond 350 AU suggest that the dust
in AB Aur is less evolved than in other T Tauri disks.

\section{Origin of the disturbances}
The AB Aur environment, although dominated by a rotating disk, is clearly highly disturbed and far
from the quasi-equilibrium Keplerian stage encountered around other young stars. We discuss here
some mechanisms which could lead to these peculiarities.

\subsection{Multiplicity ?}
Ring-like structures are easily formed by tidal truncations in binary systems, as illustrated by
the case of GG Tau \citep{Dutrey_etal1994,Guilloteau_etal1999}.  AB Aur has never been reported as
having a companion. Limits on any companion mass can be derived from the literature. For example,
using the brightness profile published by \citep{Fukagawa_etal2004} in the H band, and the
evolution models from \citet{Baraffe_etal1998, Baraffe_etal2002}, we find a (conservative,
especially at large radii) upper limit on any coeval companion of less than $0.02 \msun$ in the
field of view of the Subaru image. This upper limit on the mass holds for $120 < r < 1500$~AU. For
closer objects, the limits are more difficult to quantify. \citet{Pirzkal_etal1997} give an upper
limit of $0.25 \msun$ down to 0.4$''$ (60 AU). A stronger limit can be derived from
\citet{Leinert_etal1994} in the range $0.07 - 1.0''$ (10 to 140 AU), as their speckle observations
show that the flux of any companion in K band could not exceed 1 to 3 \% of the flux of AB Aur, for
this range of separations. This gives an upper limit of about 5.2 -- 6 to the K magnitude of a
possible companion, which translates into a mass limit of 0.05 -- 0.3 $\msun$ for ages between 1
and 10 Myr. Unfortunately, a companion of mass $M_c \ll M_*$ on an orbit with semi-major axis $a$
could create a gap of half-width $w$ a few times the Hill's radius
\begin{equation}
 w \simeq F a \sqrt[3]{\frac{M_c}{3 M_*}}
\end{equation}
where $F$ is an empirical factor of order 2 -- 3 \citep{Morbidelli_2002}. If viscosity is not
negligible, the conditions for gap creation and the resulting gap width are different. Using an
$\alpha$ prescription for the viscosity, \citet{takeuchi_etal1996} indicate a gap half width
\begin{equation}
   w \simeq 1.3 a A^{1/3}
\end{equation}
where $A$ is  the ratio of strength of tidal effects to viscous effects
\begin{equation}
  A  = \left(\frac{M_c}{ M_*}\right)^2 \frac {1}{3 \alpha (h/r)^2}
\end{equation}
Using $h/r \simeq 0.1$ as derived from our measurements, and $\alpha = 0.01$, we find $A > 1$, for
$M_c > 0.05 \msun$. Hence, viscosity is unlikely to affect our conclusions. Accordingly, even a
$0.05 \msun$ companion orbiting AB Aur around 40 AU could conceivably evacuate the inner hole.
Whether it could sustain the non Keplerian velocities requires detailed studies. Higher sensitivity
optical observations are required to setup more stringent limits on the existence of any companion.

\subsection{A gravitationally unstable disk ?}
Table \ref{tab:abaur} shows that the total mass of the system made by the disk and the envelope is
not important enough to be self-gravitating. This can be better assessed using Toomre's $Q$
criterium:
\begin{equation}
    Q = \frac{c_s \Omega}{\pi G \Sigma}~~<~~1
\end{equation}
where $c_s$ is the sound speed, and $\Omega$ the angular momentum. With our power laws, $Q$ is also
a power law of the radius $ Q(r) = Q_0 (r/R_0)^{-t}$ with
\begin{eqnarray}
    Q_0 & = & \frac{1}{\pi G} \sqrt{\frac{k}{m}} \frac{ \sqrt{T_0} \mathrm{v}_0}{\Sigma_0 R_0}  \\
    t & = & 1 + q/2 + \mathrm{v} - p
\end{eqnarray}
Using the values from Table \ref{tab:abaur}, we find $Q_0 \simeq 11$ at 100 AU and $t=-0.7 \pm
0.2$, well above the stability limit.  To have the disk massive enough for instability would only
be possible if the CO depletion was high ($> 20$) and the dust opacity 10 times lower than we
assumed. Both possibilities are highly unlikely as discussed in Section \ref{sec:mass}. For
example, lowering the dust opacity by a factor 10 would require a value of $\beta = 3.9$, assuming
the absorption coefficient at $10^{12}$ Hz remains constant. Furthermore, $t$ being negative, the
stability increases with distance from the star.
\subsection{Possible encounters}
Another possible explanation is a past encounter with a star. We have been looking in the SIMBAD
database for all stars in a radius of 30 arc-minutes centered around AB Aur. Of a total of 31 stars
referenced, 16 of them possess proper motion measurements. 9 come from the survey of
\citet{Jones_Herbig1979}. 4 come from the Tycho Reference Catalog \citep{Roeser_Bastian1988}, and 3
Hipparcos measurements were available (4 all together with AB Aur) - \citep{Perryman_etal1997}. A
very simple analysis shows that at least 2 of them could have encountered AB Aurigae in the past.
JH433 could have encountered AB Aur at any time older than 35 000 years, whereas there could have
been an encounter between AB Aur and RW Aur, some 500 000 years ago. We should also mention that
according to the early measurement of Jones and Herbig, SU Aur could have met AB Aur is a close
past, but that Hipparcos measurements rule out this possibility.

\subsection{A young disk ?}
Rather than being affected by multiplicity, the situation may suggest instead that AB Aur is
surrounded by a young circumstellar disk which has not yet relaxed to the Keplerian stage.

In the formation of a star + disk from the collapse of a rotating interstellar cloud, the low
specific angular momentum directly accretes to form the star, but the higher angular momentum
accretes through a disk. \citet{Cassen_Moosman1981} have pointed out that, because of the
projection of the momentum along the disk axis, the accreting material always have lower angular
momentum than the Keplerian value where the accreting material trajectories cross the disk.
However, when a complete history of the accretion is incorporated, \citet{Stahler_etal1994} have
shown that the disk evolution, which occurs because of the drag force induced by the accreting
material, results in a semi-steady state situation in which the outer disk is super Keplerian. Both
studies conclude that a massive ring should be formed. \citet{Stahler_etal1994} show that this ring
accumulates mass, and acts as a temporary storage for the excess angular momentum brought by the
accreting gas.

It is tempting to identify this ring with the enhanced density region which is seen in continuum
and line at about 150~AU from the star. If the inclination of the disk is low, $\simeq 23^\circ$,
the measured rotation velocities exceed the Keplerian speed at all distances above 100 AU from the
star (our measurements are insensitive to the exponent of the rotation velocity within the inner
100 AU). The apparent spiral arms could be the result of instabilities which necessarily occur in
such a situation, since the ring is a transient phenomenon.
A possible difficulty in this interpretation resides in the timescales over which the ring and
supra-Keplerian velocities persist. Accretion on the disk only occurs over the infall timescale,
which remains to first order similar to the free-fall time $t_\mathrm{ff}$, i.e. a few $10^4$ years
at most. The age of AB Aur, $t_\mathrm{AB\_Aur}$ is estimated to range between 1 and 4 Myr. 4 Myr
is cited by \cite{DeWarf_etal2003}, on the basis of a coevality with the nearby star SU Aur, which
has similar high proper motions to AB Aur. Lower ages are usually quoted by other authors, AB Aur
being among the youngest known HAeBe star. However, AB Aur is already relatively hot, being of
spectral type A0 V, so that an extremely young age, a few 10$^5$ years, appears to be ruled out
from the currently available evolutionary tracks. The main accretion phase on the disk should thus
likely be finished.

However, the longest timescale over which the circumstellar disk evolves is the viscous timescale,
\begin{equation}
  t_\mathrm{vis} (r) \approx \frac{r^2}{\nu_\mathrm{turb}} = \frac{r^2}{\alpha c_s H(r)}
\end{equation}
in the $\alpha$ prescription of the turbulence. We find
\begin{equation}
  t_\mathrm{vis} = \frac{r \mathrm{v}(r) \mu}{\alpha k T(r)} = t_\mathrm{vis}(100~\mathrm{AU})
  (r/100~\mathrm{AU})^{1-\mathrm{v}+q}
\end{equation}
Using $\alpha = 0.01$, and the temperature and velocity from Table 1, we find $t_\mathrm{vis} \simeq 10^6$~yr at
100 AU, which is comparable to the estimated age of AB Aur. It is thus conceivable that the AB Aur disk has not
relaxed to a purely Keplerian stage and stills exhibit a disturbed kinematic and density pattern. However, it is
important to mention that the viscous timescale strongly depends on the assumed viscosity parameter $\alpha$,
but little on the stellar properties. As a consequence, the above value is also applicable to T Tauri disks
which exhibit clear Keplerian signatures, like those around DM Tau, LkCa 15 or GM Aur. Unless the viscosity is
different in both type of objects, this suggests that AB Aur is younger.
\section{Summary}
We have shown from observations of dust and 4 lines from the CO isotopologues that AB Aur is
surrounded by a flattened, non-circularly symmetric disk, in rotation around the star. As expected
from the high stellar luminosity, the disk is warm, and CO is not significantly depleted. The
determination of the disk inclination is biased by the non-circularity of the brightness
distribution, but the rotation has clear non Keplerian characteristics. No completely satisfactory
explanation exists for these unusual characteristics. The disk does not appear to be massive enough
to be self-gravitating. Upper limits on any companion are not stringent enough to conclude. No
evidence for infall motions could be detected. It is tempting to see AB Aur as an example of a very
young object in which traces of the accretion from a rotating envelope onto the disk surface have
not yet been erased by the dynamical and viscous evolution of the disk. An independent indication
of youth comes from the dust properties, since the $\beta$ index may indicate that dust grains have
not evolved in the AB Aur disk as much as in other proto-planetary disks.
\begin{acknowledgements}
We acknowledge all the Plateau de Bure IRAM staff for performing the observations. We thank
J.M.Hur\'e and H.Beust for fruitful discussions on the physics of disks. The HST image was kindly
provided by C.Grady. This research has made use of the SIMBAD database, operated at CDS,
Strasbourg, France. We acknowledge financial support from the ``Programme National de Physique
Chimie du Milieu Interstellaire'' PCMI coordinated by INSU/CNRS.
\end{acknowledgements}
\bibliography{mybib}
\bibliographystyle{aa}
\end{document}